\pdfoutput=1
\documentclass[5p,10pt,sort&compress]{elsarticle}
\usepackage[english]{babel}
\usepackage{graphicx}
\usepackage[utf8]{inputenc}
\usepackage[T1]{fontenc}
\usepackage{fixltx2e}
\usepackage{url}
\usepackage{amsmath,amssymb}
\usepackage{xcolor}
\definecolor{sred}{rgb}{0.7,0,0.1}
\usepackage[colorlinks,citecolor=sred,filecolor=sred,linkcolor=sred,urlcolor=sred]{hyperref}

\newcommand{\figdir}{.}

\begin{document}

\title{Renewable build-up pathways for the US:\\ Generation costs are not system costs} 

\author[fias,stfd]{Sarah Becker\corref{cor}}
\ead{becker@fias.uni-frankfurt.de}

\author[stfd]{Bethany~A. Frew}

\author[aue,stfd]{Gorm~B. Andresen}

\author[stfd]{Mark~Z. Jacobson}

\author[fias]{Stefan Schramm}

\author[aue,aum]{Martin Greiner}

\cortext[cor]{Corresponding author}

\address[fias]{
  Frankfurt Institute for Advanced Studies,
  Goethe-Universit\"at, 
  60438~Frankfurt am Main, Germany
}

\address[stfd]{
  Department of Civil and Environmental Engineering, 
  Stanford University,
  Stanford, CA, USA
}


\address[aue]{
  Department of Engineering,
  Aarhus University, 
  8200~Aarhus~N, Denmark
}

\address[aum]{
  Department of Mathematics,
  Aarhus University, 
  8000~Aarhus~C, Denmark
}

\begin{abstract}
  The transition to a future electricity system based primarily on wind and
  solar PV is examined for all regions in the contiguous US. We present
  optimized pathways for the build-up of wind and solar power for least backup
  energy needs as well as for least cost obtained with a simplified, lightweight
  model based on long-term high resolution weather-determined generation data.
  In the absence of storage, the pathway which achieves the best match of
  generation and load, thus resulting in the least backup energy requirements,
  generally favors a combination of both technologies, with a wind/solar PV
  energy mix of about 80/20 in a fully renewable scenario. The least cost
  development is seen to start with $100\,\%$ of the technology with the lowest
  average generation costs first, but with increasing renewable installations,
  economically unfavorable excess generation pushes it toward the minimal backup
  pathway. Surplus generation and the entailed costs can be reduced
  significantly by combining wind and solar power, and/or absorbing
  excess generation, for example with storage or transmission, or by coupling the
  electricity system to other energy sectors.
\end{abstract}

\begin{keyword}
{energy system design}, {large-scale integration of renewable power generation},
{renewable power generation}, {optimal mix of wind and solar PV}, {levelized
cost of electricity}
\end{keyword}

\maketitle

\section{Introduction}

We investigate highly renewable electricity scenarios for the contiguous US. In
this paper, the main focus is placed on the optimization of the mix of wind and
solar PV power during the renewable build-up. While numerous studies investigate
regional or nationwide fully renewable power systems \citep{Budischak13,
Nelson12, Williams12, hart11, NREL_re_futures, ecf2050, energynautics}, they
usually focus on detailed single scenarios or pathways and/or only cost-optimal
installations. Here, a simplified and computationally lightweight description
based on high-resolution wind, solar PV, and load data is used to survey a large
number of possible renewable scenarios and derive systematic insights from the
spatio-temporal characteristics of the generation-load mismatch.

In our model of the electricity system, the supply is largely reliant on the
variable renewable energy sources wind and solar PV power, which we abbreviate
as VRES. Concentrated solar power (CSP) is not implemented yet. The rest of the
electricity generation is assumed to be dispatchable, and it is implied that it
is used to cover the residual demand that remains after VRES generation has been
subtracted from the load. From this point of view, the dispatchable part of the
power system will be referred to as the backup system, and correspondingly, the
energy from this system will be termed backup energy. Examples for backup power
plants in a fully renewable setting are hydroelectric power, geothermal power,
and to some extent CSP with thermal storage. In general, any other form of
dispatchable generation can be used. The share of VRES in the system is measured
as gross share, i.e.\ the total VRES generation divided by the total load. Due
to temporal mismatches in generation and load, the VRES net share, i.e.\ the
amount of VRE actually consumed in the electricity system at the time of their
generation is generally lower. Even in a system with a VRES gross share of
$100\,\%$, the load will partly be covered from backup. This renders
contributions from dispatchable renewable sources crucial to a fully renewable
system. 

\begin{table}
  \centering
  \caption{Currently (2012) installed renewable capacities in the US, as
    reported by the US Department of Energy \citep{DOE12}. The reference gives
    the installations on a state basis, and they have been aggregated into FERC
    regions using the following approximations (FERC borders and state borders
    often, but not always, coincide, cf.\ Fig.~\ref{fig:fercs}): AllCA --
    California; ERCOT -- Texas; ISONE -- Maine, New Hampshire, Vermont,
    Massachusetts, Connecticut, Rhode Island; MISO -- North Dakota, South
    Dakota, Minnesota, Iowa, Missouri, Michigan, Wisconsin, Illinois, Indiana;
    NW -- Washington, Oregon, Idaho, Montana, Wyoming, Nevada, Utah; NYISO --
    New York; PJM -- Ohio, West Virginia, Virginia, Maryland, Delaware,
    Pennsylvania, New Jersey; SE -- Arkansas, Kentucky, Tennessee, Mississippi,
    Alabama, Georgia, North Carolina, South Carolina, Florida; SPP -- Nebraska,
    Kansas, Oklahoma, Louisiana; SW -- Arizona, New Mexico, Colorado.
    Abbreviations are Geo.\ -- Geothermal, Bm.\ Biomass. All installed
    capacities are given in GW.
  }
  \label{tab:nowinstall}
  \begin{tabular}{lrrrrrr}
    \hline
    FERC  & Wind  & PV   & CSP   & Geo. & Bm. & Hydro\\
    \hline
    \hline        
    AllCA & 5.54  & 2.56 &  0.36 &  2.7 & 1.3 & 10.1 \\
    ERCOT & 12.21 & 0.14 &  0.00 &  0.0 & 0.5 & 0.7  \\
    ISONE & 0.83  & 0.29 &  0.00 &  0.0 & 1.7 & 1.9  \\
    MISO  & 17.79 & 0.12 &  0.00 &  0.0 & 1.6 & 4.1  \\
    NW    & 9.47  & 0.44 &  0.06 &  0.6 & 0.9 & 36.2 \\
    NYISO & 1.64  & 0.18 &  0.00 &  0.0 & 0.5 & 4.7  \\
    PJM   & 2.48  & 1.38 &  0.00 &  0.0 & 1.9 & 2.6  \\
    SE    & 0.03  & 0.40 &  0.08 &  0.0 & 4.6 & 13.2 \\
    SPP   & 6.31  & 0.02 &  0.00 &  0.0 & 0.5 & 1.3  \\
    SW    & 3.32  & 1.61 &  0.04 &  0.0 & 0.1 & 3.4  \\
    total & 59.62 & 7.13 &  0.55 &  3.3 & 13.4 & 78.16 \\
    \hline
  \end{tabular}
\end{table}
\begin{table*}
  \centering
  \caption{Estimated maximal installed wind and solar PV power capacities, as
    well as average and maximal installation densities $\rho$. These would occur
    if a VRES gross share of $100\,\%$ was attained with wind resp.\ solar PV
    only.  Capacity values are based on 2006/07 load averages.
  }
  \label{tab:maxinstall}
  \begin{tabular}{lrrrrrr}
    \hline
    FERC   & $\rm \frac{max.~wind~cap.}{GW}$ & $\rm \frac{max.~PV~cap.}{GW}$
           & $\frac{\rho^{\rm wind}_{\rm avg}}{\rm (MW/km^2)}$ 
           & $\frac{\rho^{\rm wind}_{\max}}{\rm (MW/km^2)}$ 
           & $\frac{\rho^{\rm PV}_{\rm avg}}{\rm (MW/km^2)}$
           & $\frac{\rho^{\rm PV}_{\max}}{\rm (MW/km^2)}$ \\
    \hline        
    \hline        
    AllCA & 130.4 & 169.1 & 0.29 & 8.2  & 0.41 &0.67 \\
    ERCOT & 149.2 & 201.5 & 0.30 & 23.2 & 0.40 &0.48 \\
    ISONE & 35.6  & 98.1  & 0.11 & 3.6  & 0.51 &0.58 \\
    MISO  & 222.2 & 406.4 & 0.13 & 2.2  & 0.26 &0.31 \\
    NW    & 105.5 & 140.7 & 0.06 & 2.5  & 0.08 &0.12 \\
    NYISO & 51.6  & 126.5 & 0.29 & 6.8  & 0.95 &1.19 \\
    PJM   & 220.8 & 523.5 & 0.37 & 8.9  & 1.04 &1.21 \\
    SE    & 530.1 & 755.1 & 0.46 & 39.6 & 0.65 &0.77 \\
    SPP   & 74.7  & 129.5 & 0.09 & 2.4  & 0.15 &0.18 \\
    SW    & 80.8  & 120.2 & 0.08 & 1.9  & 0.12 &0.15 \\
    \hline
  \end{tabular}
\end{table*}
To get an impression of the dimensions of the installations, current and
extrapolated renewable installations are shown in Tabs.~\ref{tab:nowinstall} and
\ref{tab:maxinstall}. Currently, most of the renewable power capacity is hydro
power with a total of almost $80\,\rm GW$ in 2012, closely followed by wind with
a total of close to $60\,\rm GW$. Other technologies are dwarfed in comparison,
but solar PV power has seen high growth rates over the past years \citep{DOE12}.
The largest future renewable potentials are projected to lie in wind and solar
power and are claimed to be sufficient to cover the world energy demand
\citep{jacobson11, delucchi11}, so we concentrate on these. When extrapolating
wind and solar capacities to the point where they reach a gross share of
$100\,\%$, maximal total capacities as given in the first two columns of
Tab.~\ref{tab:maxinstall} result. These capacities are theoretical estimates for
the total installed capacity in each FERC region in a hypothetical setting where
wind power (first column) resp.\ solar PV power (second column) alone produces
on average what is consumed. It is seen that even in this upper bound case,
average installation densities in each FERC region (third and fifth column of
Tab.~\ref{tab:maxinstall}) remain feasible in all regions. Only the most
concentrated wind sites in ERCOT and SE, at which maximal wind installation
densities of $23.2\,\rm MW/km^2$ resp.\ $39.6\,\rm MW/km^2$ occur (cf.\ fourth
column of Tab.~\ref{tab:maxinstall}), will need to be redistributed to
neighboring grid cells, which should not be a problem viewed in the light of the
low average wind installation densities. Solar installation densities remain
moderate even at the most concentrated sites, cf.\ the sixth column of
Tab.~\ref{tab:maxinstall}.

We make a couple of simplifying assumptions: No ramping limits are imposed on
the backup system, entailing no surplus generation from backup plants. The
slopes in both the load time series and the residual load are given in
Tab.~\ref{tab:slopes}. Column 1 gives the average slope in the load (taking no
renewable production into account), column 2 is the maximal slope of the load,
and column 3 and 4 are the average and maximal slopes of the residual load for
the case of $100\,\%$ wind and solar gross share with a backup energy minimizing
wind/solar mix, see Sec.~\ref{sec:damb} for details. All slopes are normalized
by the average load. It is seen that while the average slope does not increase
much, extreme slopes rise from around $15\,\%$ of the average load to
$70$-$100\,\%$ of the average load within one hour, indicating the need for a
more flexible backup system.
\begin{table}
  \centering
  \caption{Slopes of the part of the electrical load to be covered by the backup
    system, for the case of no wind and solar production (first two columns,
    $m^{(L)}$), and the case of $100\,\%$ wind and solar gross share (third and
    fourth column, $m^{(R)}$). The wind/solar ratio in the latter case is
    determined as the backup energy minimizing mix, cf.\ Sec.~\ref{sec:damb}.
    Shown are average (subscript "avg") as well as maximal (subscript "max")
    values. All slopes are normalized by the mean load in their respective FERC
    region.
  }
  \label{tab:slopes}
  \begin{tabular}{lrrrr}
    \hline
    FERC  & $m^{(L)}_\text{avg}$ & $m^{(L)}_\text{max}$ & $m^{(R)}_\text{avg}$ &
    $m^{(R)}_\text{max}$\\
    \hline
    \hline        
    AllCA & 0.04 & 0.16 & 0.05 & 0.92 \\
    ERCOT & 0.04 & 0.19 & 0.06 & 0.87 \\
    ISONE & 0.04 & 0.17 & 0.06 & 0.81 \\
    MISO & 0.03 & 0.13 & 0.05 & 0.74 \\
    NW & 0.03 & 0.14 & 0.05 & 0.63 \\
    NYISO & 0.03 & 0.15 & 0.06 & 0.72 \\
    PJM & 0.03 & 0.14 & 0.05 & 0.75 \\
    SE & 0.04 & 0.14 & 0.05 & 0.72 \\
    SPP & 0.03 & 0.21 & 0.05 & 0.94 \\
    SW & 0.03 & 0.14 & 0.05 & 0.99 \\
    \hline
  \end{tabular}
\end{table}

Additional measures of matching VRES generation and demand, such as storage or
demand-side management, are not treated explicitly. Likewise, potential future
changes in load characteristics or load flexibility, which may arise e.g.\ due
to electric cars, are not directly taken into account. Whenever VRES generation
exceeds the demand, surplus energy production occurs. This surplus is initially
assumed to be of no value in our model. The effect of surplus energy being sold,
possibly at a lower price, to storage, transmission, or to cover other (partly)
flexible demand like electric vehicle charging or synthetic fuel production, is
investigated later in this paper. Additionally, sensitivities to different price
assumptions are examined.

The core model has been developed and applied to obtain optimal mixes in fully
renewable energy systems as well as potential transmission grid extensions by
\citet{sarahUS}. Here, it is applied to different build-up pathways toward a
fully renewable electricity supply.

This paper is starts with a short description of the underlying data and
methodology in Sec.~\ref{sec:dam}. Subsequently, the resulting US build-up
pathways and their sensitivities to cost assumptions and surplus usage are
presented in Sec.~\ref{sec:res}. Sec.~\ref{sec:concl} summarizes the main
findings and concludes the paper.

\section{Data and methodology}
\label{sec:dam}

\subsection{Load and generation data}
\label{sec:dama}

\begin{figure}
  \centering
  \includegraphics[width=0.49\textwidth]{\figdir/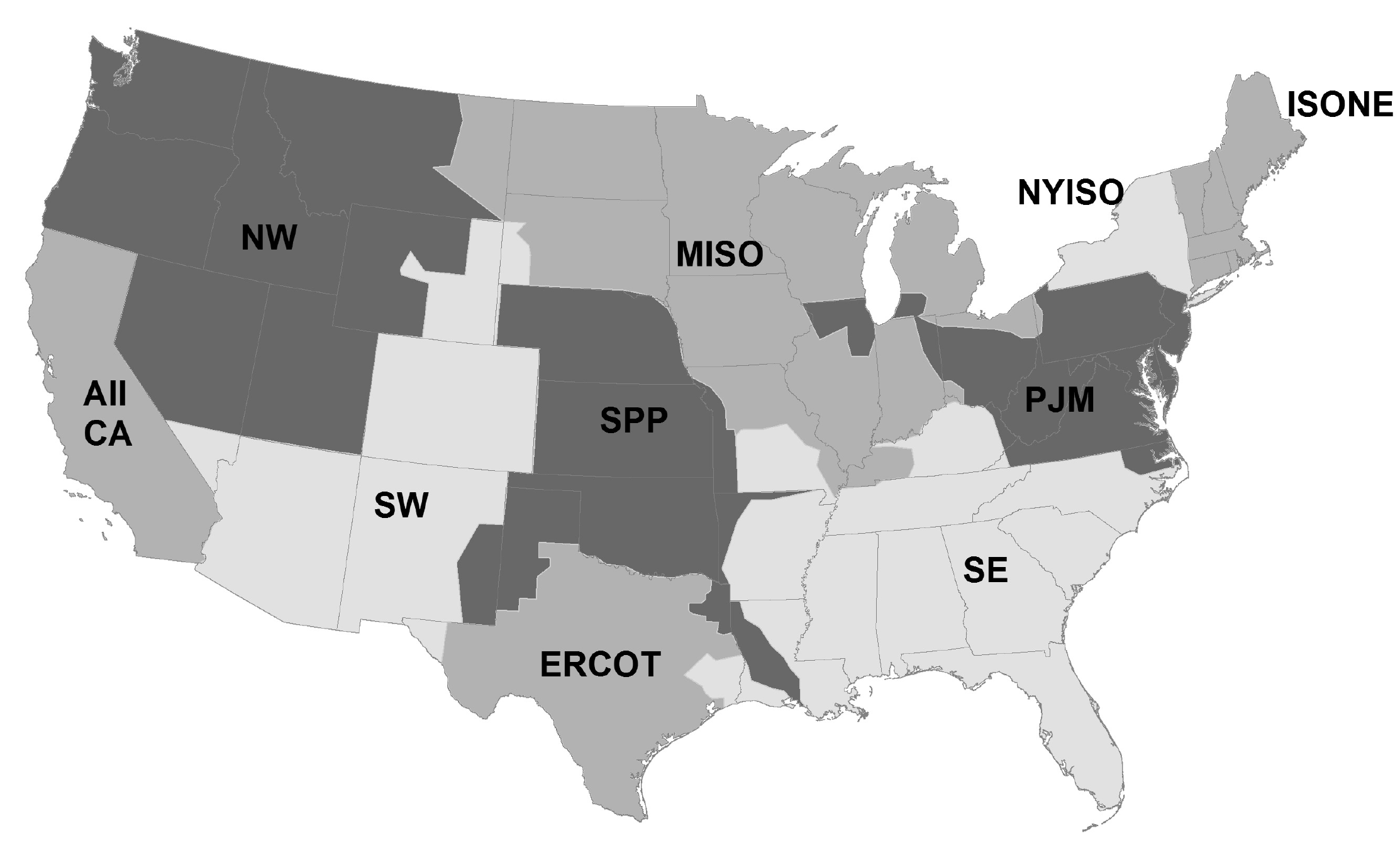}
  \caption{Federal Electricity Regulatory Council (FERC) regions of the
    contiguous US, based on \cite{bethany}.
  }
  \label{fig:fercs}
\end{figure}
The analysis is based on weather data for 32 years with one hour time steps and
$30\times30\,\rm km^2$ grid cells, covering the time span 1979-2010, from the
NCEP (National Centers for Environmental Prediction) Climate Forecast System
Reanalysis \citep{saha}. They were converted to wind and solar PV generation data
as described in by~\cite{sarahUS, REAtlas, anders}.  Wind capacity layouts
were chosen similarly to those used to produce the National Renewable Energy
Laboratory (NREL) wind datasets \citep{NREL_wwind, NREL_ewind}, while solar PV
capacity was distributed according to the potential generation in each grid
cell. Solar panels with a nameplate capacity of $156\,\rm kW$ fixed in southward
direction at a tilt equal to the latitude were assumed. This tilt implies that
the panel orientation is optimal for the average solar noon position. In our
data, solar capacity factors between $15\,\%$ (in ISONE and NYISO) and $20\,\%$
(in California and SW) are observed. $3\,\rm MW$ wind turbines with a hub height
of $80\,\rm m$ onshore and $7\,\rm MW$ at $100\,\rm m$ hub height offshore were
assumed, yielding average capacity factors between $23\,\%$ in SE and $42\,\%$
in ISONE, see Tab.~\ref{tab:cost}. Power generation from each grid cell was
aggregated to Federal Electricity Regulatory Council (FERC) region level. See
Fig.~\ref{fig:fercs} for a map of the contiguous US FERC regions.  Details of
the data processing can be found in \cite{sarahUS}.

Historical load data for the years 2006-2007 were compiled for each FERC region
in \cite{bethany}.  Where necessary, load data were extended by repetition
to cover the 32-year simulation period.

The aggregation of wind and solar PV generation as well as load implies that no
FERC-region-internal bottlenecks are present in the transmission grid. It is
indeed likely that in a highly renewable electricity system, the regional
transmission grids will be reinforced, because of the beneficial effects of
aggregation on smoothing wind and solar PV output, well documented in the
scientific literature, e.g.\ \cite{archer, holttinen05, sinden07, wiemken01,
mills10, Widen:2011ys, Kempton:2010oq}. Inter-FERC-region transmission has the
potential to smooth VRES generation even further \citep{delucchi11,sarahUS}, but
is initially not incorporated into the model.

Central to our research is the mismatch $\Delta_n$ between load $L_n$ and
generation $G_n^{\rm S}$, $G_n^{\rm W}$ from solar PV and wind, respectively, in
FERC region $n$.
\begin{align}
  \Delta_n(t) = \gamma_n \,\langle L_n\rangle 
  \left[(1-\alpha_n^{\rm W} )G_n^{\rm S}(t) + \alpha_n^{\rm W}G_n^{\rm W}(t) \right]
  -L_n(t)
  \label{eq:mism}
\end{align}
In this expression, wind and solar generation are understood to be normalized to
an average of one, and then scaled with the mean load $\langle L_n\rangle$ to a
given gross share $\gamma_n$ of the load. The relative share of wind in the VRE
generation is denoted $\alpha_n^{\rm W}$, the corresponding relative share of
solar PV is $(1-\alpha_n^{\rm W} )$. 

Note that the VRES gross share $\gamma_n$ is the ratio between the average VRES
production and the average load, not to be confused with the share of VRES
electricity in the total consumption, the VRES net share. This is due to the
fluctuating nature of VRES generation, which especially for $\gamma_n > 50\,\%$
leads to surplus VRES production that does not contribute to covering the
electric load. $\gamma_n$ is an upper bound on the percentage of VRES in the
electricity mix.

\subsection{Backup energy-minimal mix}
\label{sec:damb}

Here, the only concern is to keep the need for backup energy, which is
calculated as the sum of negative mismatches throughout all time steps, as small
as possible. In other words, the sum of the negative parts (denoted $(.)_-$) of
the mismatch in Eq.~\eqref{eq:mism} is minimized as a function of $\alpha_n^{\rm
W}$:
\begin{align}
  \min_{\alpha_n^{\rm W}} \sum_{t} \left( \Delta_n(t)\right)_-
  \label{eq:balmin}
\end{align}
The backup energy minimization leading to Fig.~\ref{fig:pathex} is performed
independently for different VRE gross shares $\gamma_n$. Since in our modeling,
the VRE gross share $\gamma_n$ and hence the total energy produced from VRES is
fixed, least backup energy needs are equivalent to least surplus VRES
generation. In other words, when minimizing the need for backup energy from
dispatchable sources, the VRES surplus energy is minimized at the same time.

When the VRES gross share $\gamma_n$ is less than $100\,\%$, at least a fraction
of $(1-\gamma_n)$ of the demand has to be covered by the backup system, even if
no VRES generation comes as surplus energy. The energy provided by the backup
system beyond this minimal share is termed additional backup energy, and this is
the part of the backup energy that can be reduced by a suitable choice of the
wind/solar mix.

\subsection{Regional LCOE}
\label{sec:damc}

\begin{table}
  \centering
  \caption{Relative regional levelized costs of electricity (LCOE) for the 10
    FERC regions, for solar (left) and wind (right), together with the capacity
    factor $CF_n$, resource-quality related weight factor $w_n$ (see
    Eq.~\ref{eq:w_n}), and regional material and labor cost weights $c_n$ (from
    \cite{E3report}). Cost differences due to building in easier accessible
    or more remote areas are not presently included.
  }
  \label{tab:cost}
  \begin{minipage}{0.49\textwidth}
    \centering
    \begin{tabular}{lrrrr}
      \hline
      Region & $CF_n$ & $w_n$ & $c_n$ & rel.\ $\text{LCOE}_n$ \\                               
      \hline        
      \hline                                                                                   
      AllCA & 0.20 & 0.87 & 1.04 & $89.8\,\%$ \\                                                   
      ERCOT & 0.18 & 0.98 & 0.97 & $94.0\,\%$ \\                                                   
      ISONE & 0.15 & 1.12 & 1.02 & $113.5\,\%$ \\                                                  
       MISO & 0.17 & 1.03 & 1.01 & $103.6\,\%$ \\                                                   
         NW & 0.19 & 0.91 & 1.00 & $90.4\,\%$ \\                                                      
      NYISO & 0.15 & 1.15 & 1.10 & $125.6\,\%$ \\                                                  
        PJM & 0.16 & 1.10 & 1.03 & $112.8\,\%$ \\                                                    
         SE & 0.16 & 1.05 & 0.94 & $98.2\,\%$ \\                                                      
        SPP & 0.18 & 0.94 & 0.96 & $90.0\,\%$ \\                                                     
         SW & 0.20 & 0.84 & 0.98 & $82.1\,\%$ \\                                                      
       avg. & 0.17 & 1.00 & 1.00 & $100.0\,\%$ \\    
      \hline
    \end{tabular}
  \end{minipage}
  \begin{minipage}{0.49\textwidth}
    \centering
    \begin{tabular}{lrrrr}
      \hline
      Region & $CF_n$ & $w_n$ & $c_n$ & rel.\ $\text{LCOE}_n$ \\
      \hline
      \hline        
      AllCA & 0.25 & 1.15 & 1.04 & $119.5\,\%$ \\
      ERCOT & 0.24 & 1.24 & 0.97 & $120.0\,\%$ \\
      ISONE & 0.42 & 0.70 & 1.02 & $71.0\,\%$ \\
       MISO & 0.30 & 0.97 & 1.00 & $96.8\,\%$ \\
         NW & 0.25 & 1.17 & 1.00 & $116.9\,\%$ \\
      NYISO & 0.36 & 0.80 & 1.04 & $83.6\,\%$ \\
        PJM & 0.37 & 0.80 & 1.01 & $80.5\,\%$ \\
         SE & 0.23 & 1.27 & 0.98 & $124.0\,\%$ \\
        SPP & 0.31 & 0.93 & 0.98 & $91.5\,\%$ \\
         SW & 0.30 & 0.97 & 0.99 & $96.2\,\%$ \\
       avg. & 0.30 & 1.00 & 1.00 & $100.0\,\%$ \\
      \hline
    \end{tabular}
  \end{minipage}
\end{table}
Wind and solar PV levelized costs of electricity (LCOE) are expected to vary
spatially due to different external conditions. The inhomogeneity is captured by
region-specific cost factors in our model. The main cause of deviations is the
weather-dependent capacity factor $CF_n$ for each region (indexed $n$), i.e.\
the ratio of average generated power to the maximal generator capacity.  Since
the costs of VRE plants are to a large part installation and maintenance costs
which are proportional to the total installed capacity, but almost independent
of the total power output, the costs per unit of energy are in good
approximation anti-proportional to the total generated energy. Expressed in
terms of the capacity factor, this yields a regional weight factor of 
\begin{align}
  w_n = \frac{N}{\sum_m 1/CF_m} \cdot \frac{1}{CF_n}
  \label{eq:w_n}
\end{align}
The normalization (first factor) is necessary to keep the average of the weights
at unity. $N$ is the number of regions, in this case, 10.

The second reason for variations in LCOE in the FERC regions are different labor
and material costs, which have been compiled by the US Army Corps of Engineers
\citep{cwccis}, and adapted to the problem at hand in \citet{E3report}.
These yield another factor $c_n$ of the order of one, which modifies the
regional LCOE. The accessibility of the average site in each region is another
factor that can influence installation and maintenance costs, but is not
accounted for here.  Taken together, the regional LCOE are calculated as:
\begin{align}
  \text{LCOE}_n = w_n c_n \text{LCOE}_\text{avg}
  \label{eq:regcost}
\end{align}
They are determined separately for wind and solar PV.  The capacity factor
weights $w_n$ as well as the regional cost factors $c_n$ are given in
Tab.~\ref{tab:cost}, which also shows the relative LCOE in the different FERC
regions, for solar PV and wind power separately.

As a rough guess, we assume equal wind and solar PV LCOE$_\text{avg}$ of
\$0.04/kWh. This choice is in accordance with the most recent Lazard LCOE
estimates \citep{Lazard2014}, reporting 2014 unsubsidized LCOE for wind in the
range \$0.031-\$0.087/kWh and for utility scale solar PV \$0.072-\$0.086/kWh.
Given that the renewable shares discussed in this paper represent mid- to
far-future scenarios, and the steeper historical LCOE reductions for solar PV as
compared to wind, the assumption of \$0.04/kWh for both appears reasonable. It
should be noted that our results depend only on the relative costs of wind and
solar PV (with the obvious exception of the future absolute LCOE), and so remain
valid even if LCOE see a slightly different development. Furthermore, we have
investigated the effect of LCOE changes in a sensitivity analysis
(Sec.~\ref{sec:LCOEsens}).

\subsection{LCOE-minimal mix}

For each region $n$, the local wind and solar PV LCOE resulting from the
regionalization, Eq.~\eqref{eq:regcost}, are combined into an average regional
LCOE of VRES, depending on the relative wind share in VRES, $\alpha_n^{\rm W}$:
\begin{align}
  \nonumber
  &\text{LCOE}_{0,n}(\alpha_n^{\rm W}) = \alpha_n^{\rm W} \text{LCOE}_n^{\rm W} +
  (1-\alpha_n^{\rm W})\text{LCOE}_n^{\rm S}
\end{align}
These are then modified to account for the effects of surplus production: It is
initially assumed that the surplus production has no value and thus effectively
raises LCOE by reducing the amount of usable electric energy produced, as stated
in Eq.~\ref{eq:prices1} below.
\begin{align}
  \begin{split}
    \text{LCOE}_{\text{mod.},n}(\alpha_n^{\rm W}) &=  
    \text{LCOE}_{0,n}(\alpha_n^{\rm W})\cdot \\
    &\quad \frac{E_\text{generated}(\alpha_n^{\rm W})}
    {E_\text{generated}(\alpha_n^{\rm W})-E_\text{surplus}(\alpha_n^{\rm W})}
  \end{split}
  \label{eq:prices1}
\end{align}
$E_\text{generated}$ is the total energy generated from VRES, and
$E_\text{surplus}$ is the VRES surplus energy.  Notice that the amount of
surplus energy here equals the amount of additional backup energy requirements
due to VRES fluctuations, discussed in the section above.

\section{Results}
\label{sec:res}

\subsection{Minimal backup energy pathways}

\begin{figure}
  \centering
  \includegraphics[width=0.45\textwidth]{\figdir/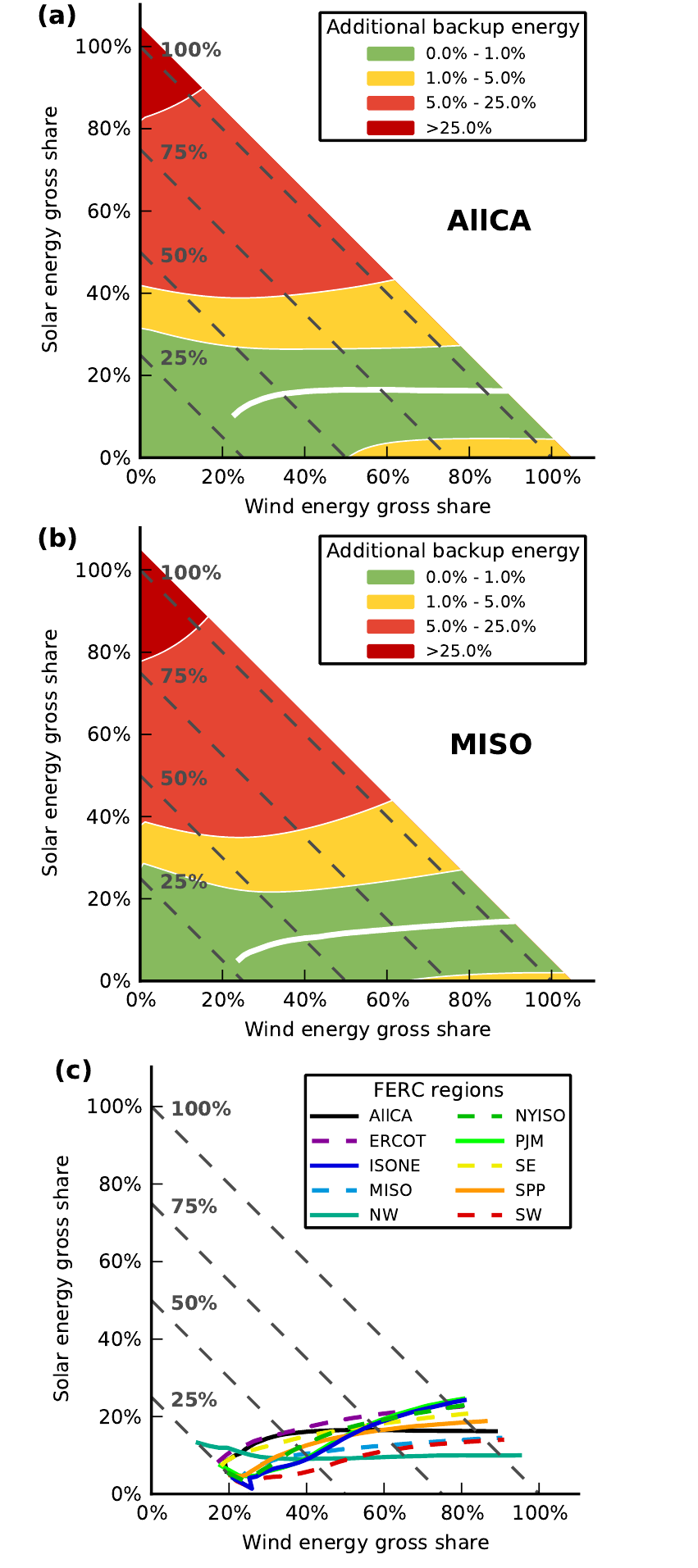}
  \caption{(a), (b): Build-up pathway from a renewable gross share of $0\,\%$ to
    $100\,\%$ for (a) California and (b) MISO that minimizes backup energy needs
    (Eq.~\eqref{eq:balmin} in Sec.~\ref{sec:dama}) during the entire renewable
    build-up. In each of these plots, the white line indicates the build-up
    pathway minimizing backup energy requirements at the later stages of the
    installation process. In the green region, backup energy is up to
    1~percentage point (pp) of the load larger than optimal. In the yellow
    region, it is up to 5~pp larger than optimal. In the light red region, it is
    up to 25~pp larger, and in the dark red region, more than 25~pp larger. The
    dark gray dashed lines indicate renewable gross shares $\gamma_n$ of
    $25\,\%$, $50\,\%$, $75\,\%$, and $100\,\%$. (c): Build-up pathways
    minimizing backup energy for all FERC regions, analogous to the white line
    in (a) and (b), starting from $25\,\%$ VRE gross share. For lower shares,
    the minimum in backup energy as a function of wind/solar mix is very
    shallow. This leads to fluctuations in the optimal mix as a function of VRE
    gross share, which are not indicative.  The optimal mix is therefore only
    shown above a VRE share of $25\,\%$.
  }
  \label{fig:pathex}
\end{figure}
\begin{figure}
  \centering
  \includegraphics[width=0.45\textwidth]{\figdir/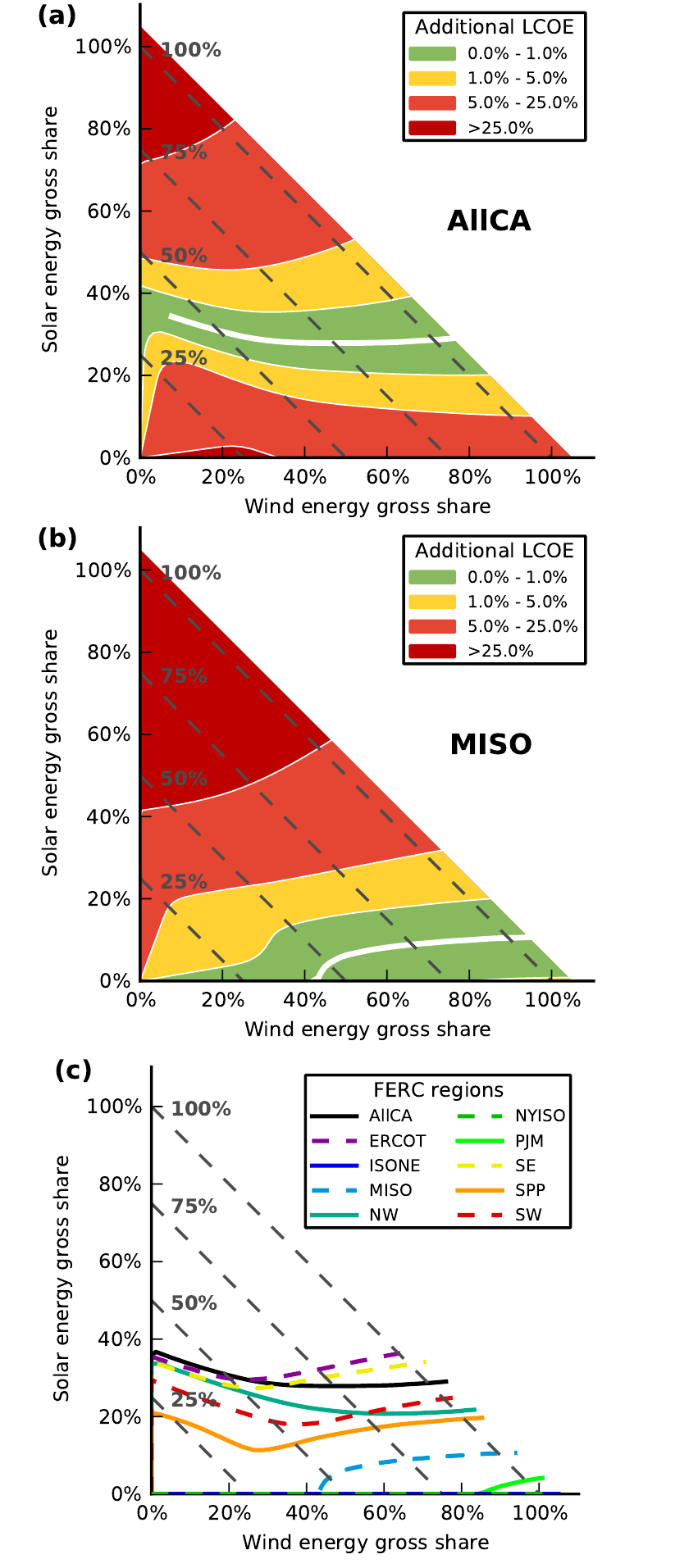}
  \caption{(a), (b): Build-up pathway from a renewable gross share of $0\,\%$ to
    $100\,\%$ for (a) California and (b) MISO that minimizes combined renewable
    LCOE (Eq.~\ref{eq:prices1}) during the entire renewable build-up. In each of
    these plots, the white line indicates the build-up pathway minimizing LCOE
    at the later stages of the installation process. In the green region, LCOE
    are up to 1~percentage point (pp) larger than optimal, in the yellow region,
    up to 5~pp, in the light red region, up to 25~pp, and in the dark red
    region, more than 25~pp larger. The dark gray dashed lines indicate 
    renewable gross shares $\gamma_n$ of $25\,\%$, $50\,\%$, $75\,\%$, and
    $100\,\%$. LCOE are assumed to be equally \$0.04/kWh for both wind and solar
    PV on average across the contiguous US, which translates into \$0.048/kWh
    for wind and \$0.036/kWh for solar PV in California and \$0.039/kWh for wind
    and \$0.041/kWh for solar PV in MISO when LCOE are regionally adjusted, see
    Eq.~\eqref{eq:regcost}.  (c): LCOE-minimal build-up pathways for all FERC
    regions, analogous to the white lines in (a) and (b).  Note that since the
    LCOE-minimal mix for ISONE and NYISO is $100\,\%$ wind during the entire
    build-up, their pathways coincide with the x-axis.
  }
  \label{fig:pathexl}
\end{figure}
\begin{figure}
  \centering
  \includegraphics[width=0.45\textwidth]{\figdir/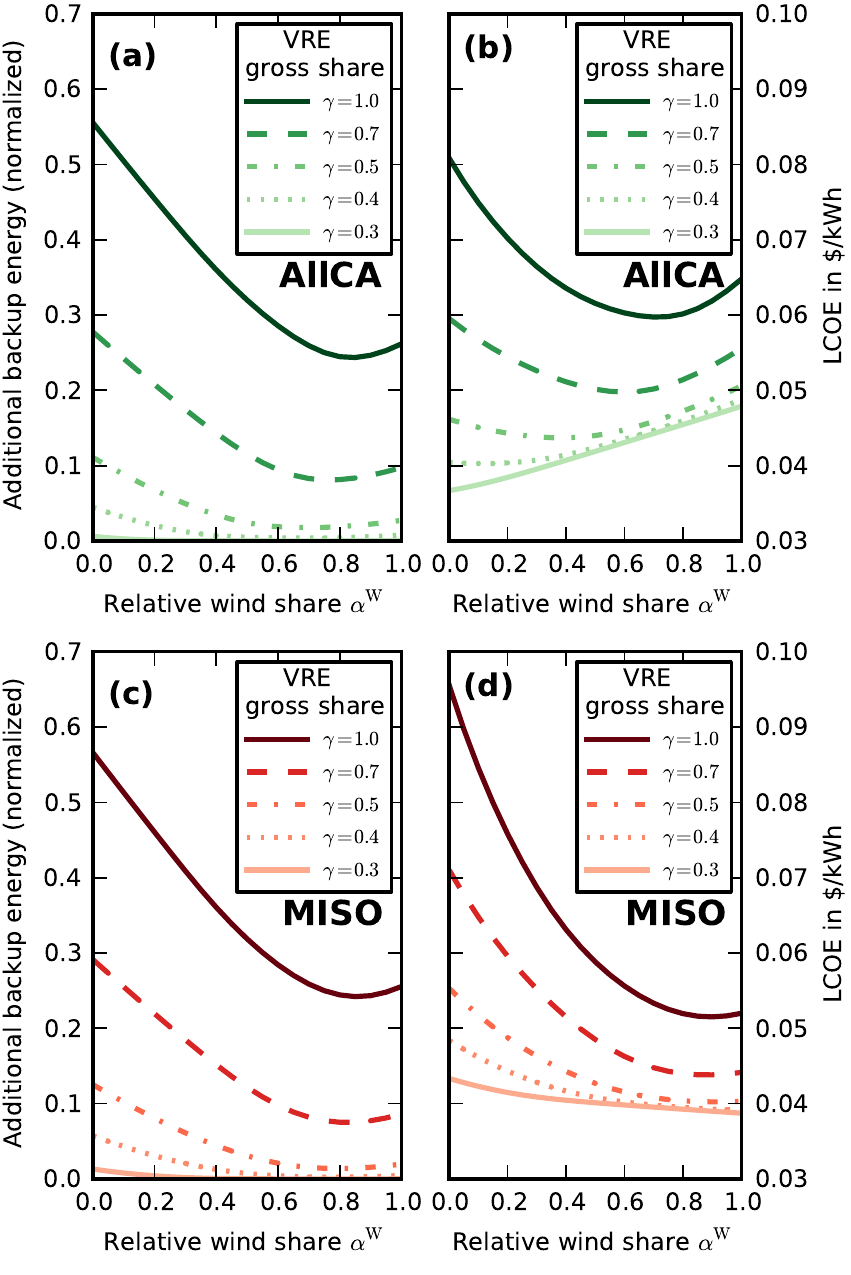}
  \caption{(a), (c): Additional backup energy, normalized by the average load,
    and (b), (d): LCOE as a function of the wind/solar mix, for different VRES
    gross shares $\gamma$ between $30\,\%$ and $100\,\%$, in the example FERC
    regions (a), (b) AllCA and (c), (d) MISO.
  }
  \label{fig:min}
\end{figure}

Backup energy minimizing build-up pathways have been calculated by optimizing
the wind/solar mix for VRE gross shares between $0\,\%$ and $100\,\%$, see
Eq.~\eqref{eq:balmin} in Sec.~\ref{sec:dama}. Detailed examples are shown in
Fig.~\ref{fig:pathex}a for California and in Fig.~\ref{fig:pathex}b for MISO.
The minimizing pathways for all other FERC regions are included in
Fig.~\ref{fig:pathex}c. Figs.~\ref{fig:pathex}a and b present the optimal
pathway (white line), along which backup energy is minimal for each given VRES
share.  Additionally, parameter combinations that lead to increasingly more
backup energy than the optimal path are indicated: In the green region, the
average backup energy requirement is less than 1~percentage point (pp) of the
average load more than optimal, in the yellow region, 5~pp, in the red region,
25~pp, and in the dark red region, more than 25~pp. The green region is seen to
successively shrink during the build-up, showing that the minimum in backup
energy becomes more and more pronounced with growing VRE share. This observation
is corroborated by Figs.~\ref{fig:min}a and c, where sections for several fixed
renewable gross shares through the backup energy needs are shown as a function
of the wind/solar mix. Only additional backup energy is included in
Figs.~\ref{fig:min}a and c, which arises due to VRE fluctuations. It is equal to
the excess of backup energy over the expected ``missing energy'' of total
electricity demand minus total VRES generation; see Sec.~\ref{sec:dama} for
details. 

In the early stage of VRE installations, until wind and solar PV cover about
$30\,\%$ of the load, the sensitivity of backup energy need with respect to the
mix of wind and solar is relatively low, because both wind and solar PV
generation hardly ever exceed the demand, so all energy they produce can be used
in the electricity system and no additional backup energy is required. Toward a
fully renewable system, the mismatch between load and generation grows. Once VRE
gross shares reach $30\,\%$ to $50\,\%$, substantial VRE surplus generation and
hence need for additional backup energy at other times occurs, which can be
minimized using the mix of wind and solar PV as a handle. During the later
stages of the development, when VRE gross shares reach more than $50\,\%$,
backup minimal mixes for all FERC regions are observed around $80\,\%$ wind and
$20\,\%$ solar PV, with a spread of about $10\,\%$ across the different FERC
regions, cf.\ Fig.~\ref{fig:pathex}c.

\subsection{Minimal LCOE pathways}

LCOE-optimized VRES build-up paths are shown in Fig.~\ref{fig:pathexl}a for the
example of California and in Fig.~\ref{fig:pathexl}b for MISO in detail, and
similar pathways in Fig.~\ref{fig:pathexl}c for all FERC regions. As above, the
white line traces the optimal path. Here, the green region indicates scenarios
in which LCOE are up to 1~pp of the average of wind and solar PV LCOE (before
the modifications of Eq.~\ref{eq:prices1}) higher than optimal. In the yellow
region, LCOE are up to 5~pp higher than optimal, in the light red region, up to
25~pp, and in the dark red region, more than 25~pp. All pathways are calculated
under the example assumption of equal country-average VRES LCOE for wind and
solar PV of \$0.040/kWh. With the cost regionalization of
Eq.~\eqref{eq:regcost}, this yields \$0.048/kWh for wind and \$0.036/kWh for
solar PV in California and \$0.039/kWh for wind and \$0.041/kWh for solar PV in
MISO. 

In contrast to the backup optimal pathways of Figs.~\ref{fig:pathex}a-c, the
LCOE optimal mix strongly favors the lower cost technology for low renewable
penetrations -- solar PV for California, wind for MISO under our example cost
assumptions. The cause for this behavior is that both can be integrated equally
well into the system, so there is no disadvantage in picking the cheaper one.
Only when surplus production and additional backup requirements become more
prominent and expensive, around VRES gross shares of $30\,\%$ to $50\,\%$, the
mix shifts toward lower backup energy requirements. This effect is further
illustrated in Figs.~\ref{fig:min}b and d, where the shift of the LCOE minimum
from least generation cost for low VRE gross shares toward least
surplus/additional backup for higher shares is clearly visible. It can be
interpreted as an indication that although in the short run it appears cheaper
to settle for the lower generation cost resource, in the long run it pays to
sustain a mixed portfolio, which is able to reduce backup energy needs and
surplus production.

It is interesting to compare the build-up pathway for California obtained here
to the results of the more detailed SWITCH model \citep{Nelson12}. In contrast to
our modeling, they assume a solar PV installation cost about twice as high as
for wind, which results in early VRES growth almost exclusively in wind.
Subsequently, solar PV costs are assumed to decrease in a steep learning curve,
dropping almost down to the cost of onshore wind at the end of their simulation
period in 2029. This leads to significant solar installations in later years.
Similar to our modeling, VRES installations start with the lowest cost
technology, which is complemented by others in the following years, as renewable
shares grow. Due to the complexity of the SWITCH model, this analogous
development cannot, however, be traced back to the same mechanism of avoiding
backup energy needs and surplus production by shifting the mix that we observed
in our model.

Note that, since LCOE are minimized for all renewable shares independently, the
optimal build-up pathway (white lines in Figs.~\ref{fig:pathexl}a and b, colored
lines in Fig.~\ref{fig:pathexl}c) sometimes traces an uninstallation or
under-usage of previously existing renewable capacity. However, the green
region, where LCOE are less than one percent larger than optimal, is broad
enough to accommodate a modified pathway that does not include uninstallation.
An analogous statement holds for the minimal backup energy pathways,
Figs.~\ref{fig:pathex}a-c.

\subsection{Usage of surplus energy}

It can be argued that no value of all occurring surplus energy is an unrealistic
assumption. If initially there was no use for surplus electricity, it would be
available cheaply. This in turn would strongly incentivize the development of
measures to make use of the surplus. A future electricity system is therefore
likely to include sources of flexibility to capture some value from surplus
generation. For example, demand-side management measures or storage systems may
be used, reducing surplus energy. Additionally, inter-FERC region transmission
leads to surplus being exported to other parts of the country, where it can be
used to replace backup energy. It was found in \cite{sarahUS} that in a
$100\,\%$ renewable scenario, unlimited transmission reduces the residual
surplus by roughly one fifth. Another option is to use surplus electricity for
heating or transportation.

\begin{figure}
  \centering
  \includegraphics[width=0.45\textwidth]{\figdir/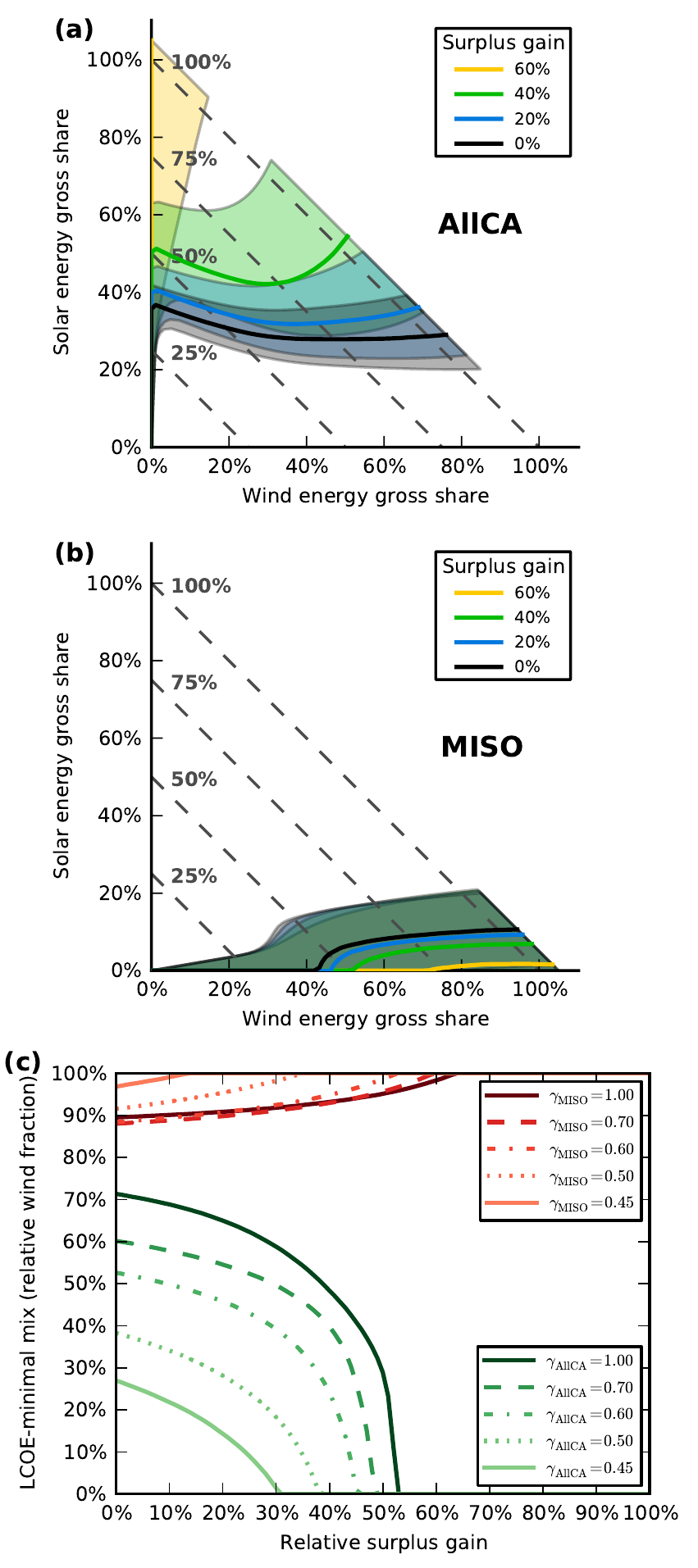}
  \caption{(a), (b): LCOE-minimizing build-up pathways if it is possible to gain
    $0\,\%$, $20\,\%$, $40\,\%$, and $60\,\%$ of the average LCOE for the
    surplus energy, for the example regions (a) AllCA and (b) MISO, for wind and
    solar LCOE of \$0.048/kWh and \$0.036/kWh (in AllCA) or \$0.039/kWh and
    \$0.041/kWh (in MISO), before accounting for lower-value surplus,
    respectively. The shadowed regions indicate the 1~pp higher LCOE wind-solar
    combinations for each surplus gain percentage. (c) shows the LCOE-minimal
    mix for different VRE gross shares as a function of the surplus gain. The
    surplus gain can be realized by selling part of the surplus for the normal
    price, or all of it for a lower than normal price, or something in between.
  }
  \label{fig:usage}
\end{figure}
To address such effects, modified LCOE-minimal pathways are investigated, where
only a fraction of the surplus is treated as not giving any gain, thus
subtracting only a fraction of the surplus energy from the total generated
energy in the denominator of Eq.~\eqref{eq:prices1}. For example, $20\,\%$ gain
on the surplus could be achieved by recovering the full LCOE of $20\,\%$ of the
surplus by selling it to some alternative consumer (e.g.\ storage, transmission,
synthetic fuel production), or by recovering part of the LCOE on a corresponding
larger fraction of the excess generation.  The results are illustrated in
Fig.~\ref{fig:usage}a and b, again for the AllCA and MISO regions. Shown are
three cases where $20\,\%$, $40\,\%$, and $60\,\%$ of the incurred LCOE are
gained from surplus energy. For AllCA, it is seen that while for the $20\,\%$
case, not much changes with respect to the no-value-surplus case depicted in
Fig.~\ref{fig:pathexl}a, already $40\,\%$ of the surplus energy's generation
costs gained means a significant shift in the LCOE-minimal path toward the
cheaper technology, in this example, solar.  However, there is still a
significant share of wind power in the $100\,\%$ LCOE-minimal mix.  This changes
beyond about $50\,\%$ of the gains on surplus energy, compare the green lines in
Fig.~\ref{fig:usage}c, when the LCOE-minimal mix shifts to solar PV all the way
to $100\,\%$ VRES gross share. 

For MISO (Fig.~\ref{fig:usage}b), the absolute shift of the pathways is smaller,
because the lower generation cost technology is wind in this case, which brings
the backup minimal mix and the LCOE minimal mix closer together from the start.
Qualitatively, however, the picture is similar: For $20\,\%$ and $40\,\%$
surplus usage, the change in the LCOE-minimal pathway is relatively small. Only
for higher surplus usage fractions, the LCOE optimal path finally shifts to
$100\,\%$ wind all the way.  This shift toward more wind with growing surplus
usage at different VRE gross shares is shown in Fig.~\ref{fig:usage}c (red
lines).  It is seen that for higher gross shares, the LCOE-minimal path reaches
$100\,\%$ wind only at a higher surplus gain than for lower gross shares.

In conclusion, a high share of the surplus energy has to be used for other goals
than satisfying the electricity demand to shift the LCOE-minimal mix back to
where it is seen on a pure generation-cost basis.

\subsection{Sensitivity to different generation costs}
\label{sec:LCOEsens}

\begin{figure}
  \centering
  \includegraphics[width=0.45\textwidth]{\figdir/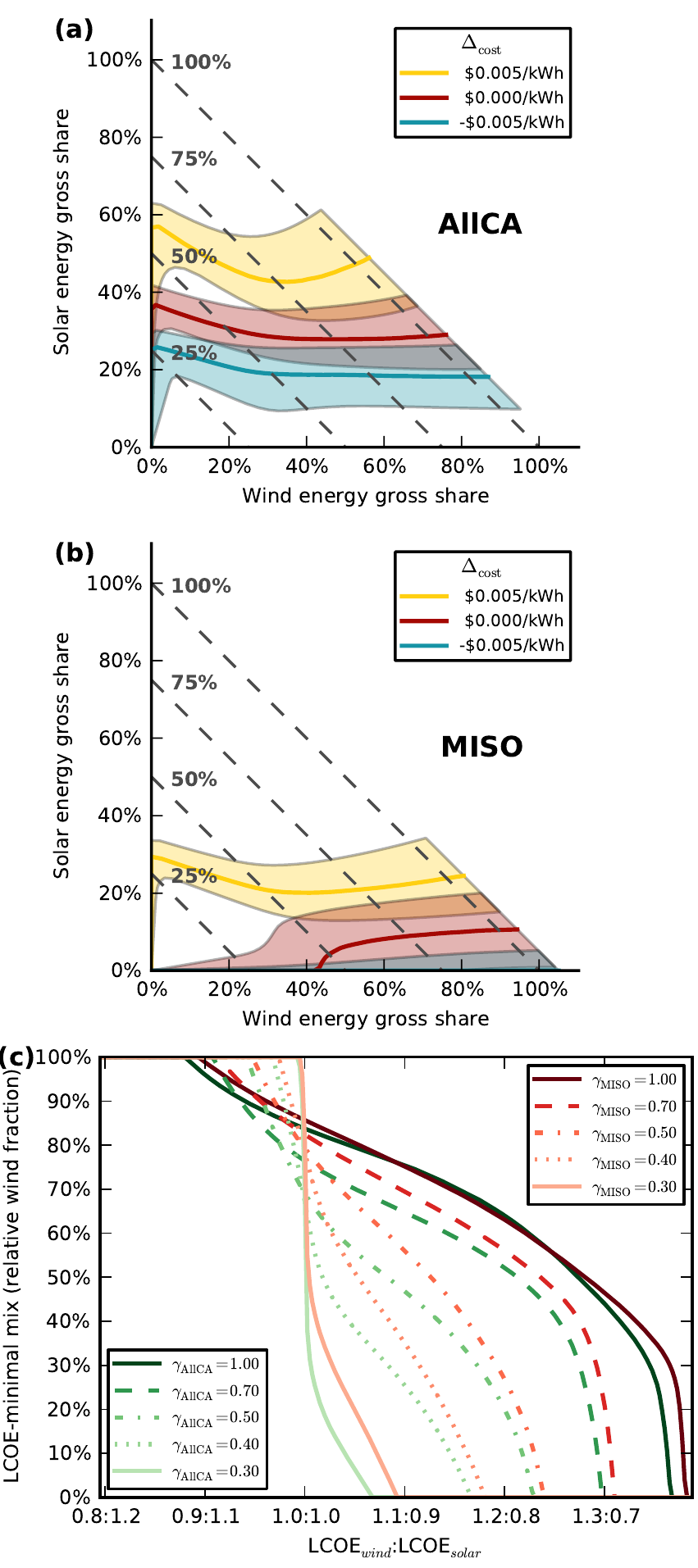}
  \caption{Cost sensitivity of the least cost build-up pathway for (a) AllCA and
    (b) MISO. Shown is the effect of cost changes on the optimal pathways,
    comparing three cases: (1) LCOE (before accounting for no-value surplus)
    remain unchanged 
    (red curves, $\Delta_\text{cost}=$ \$0.000/kWh in the legend), (2) wind LCOE
    are reduced by \$0.005/kWh and solar PV LCOE increased by \$0.005/kWh (blue
    curves, $\Delta_\text{cost}=-$\$0.005/kWh), and (3) wind LCOE increased by
    \$0.005/kWh and solar PV LCOE reduced by \$0.005/kWh ($\Delta_\text{cost}=$
    \$0.005/kWh). The shaded areas indicate the regions where LCOE are less than
    $1\,\%$ larger than optimal. (c) shows the LCOE-minimal mix as a function of
    the LCOE ratio, for five different VRE gross shares $\gamma$ between
    $30\,\%$ and $100\,\%$, for AllCA and MISO.
  }
  \label{fig:pathspan}
\end{figure}
As the overall LCOE minimization only depends on the ratio of wind and solar
LCOE (compare Eq.~\ref{eq:prices1}), the results remain unchanged as long as
wind and solar LCOE are both raised or lowered by the same percentage. An
example of this effect would be observed if learning curves with the same time
constants (start time and learning rate) were assumed for both technologies.
Such a case is therefore covered by the present work.

Figs.~\ref{fig:pathspan}a and b show what happens to Figs.~\ref{fig:pathexl}a
and b when relative cost assumptions change. They depict the LCOE-minimal
build-up of wind and solar PV if the initial LCOE are changed such that one
technology is \$0.005/kWh more expensive and the other \$0.005/kWh less
expensive. While this shifts the least-cost path toward the now cheaper
technology, it does not change the qualitative characteristics of the picture as
long as the same technology remains the cheaper one. Where solar PV has lower
generation costs, the build-up starts with $100\,\%$ solar PV, and shifts
rapidly toward more wind when additional backup needs arise. Where wind is
cheaper, the build-up starts with $100\,\%$ wind, and later gradually includes a
small fraction of solar PV.  For AllCA, cost changes are small enough such that
solar PV remains cheaper for all cost scenarios depicted in
Fig.~\ref{fig:pathspan}a. For MISO, the two different behaviors for lower solar
(yellow curve in Fig.~\ref{fig:pathspan}b) and lower wind generation costs (red
and blue curves in Fig.~\ref{fig:pathspan}b) can be seen.

The changes in these panels when larger cost changes are applied are shown in
Fig.~\ref{fig:pathspan}c, in green for AllCA and in red for MISO. If LCOE for
wind and solar PV are equal, the LCOE-minimal mix equals the backup energy
minimizing mix. For lower wind LCOE, wind quickly becomes the only generation
technology, while solar PV LCOE have to drop down to less than half of wind LCOE
to make a solar PV only mix the cheapest option.  This is due to the large
mismatch between solar generation alone and the load.  For all curves, the
sensitivity to initial LCOE becomes lower and lower (curves are less steep) with
increasing VRES gross share, because this leads to more surplus/additional
backup energy that needs to be minimized besides generation costs. The effects
of different resources on these plots are very small, as can be seen from the
comparison of AllCA and MISO -- the curves for the same gross VRES shares almost
coincide in Fig.~\ref{fig:pathspan}c.

\section{Comparison and Conclusions}
\label{sec:concl}

Fig.~\ref{fig:min}a and c show that for low VRES gross shares, surplus
production entailing additional backup energy needs hardly ever occurs, and thus
the choice of the wind/solar mix is largely irrelevant for the backup energy
minimization.  Starting from a gross share of about $30\,\%$, this changes:
Surplus production sets in, and hence backup minimization becomes more
important, leading to successively narrower minima in backup energy. The
wind/solar mix becomes more important with growing installations. In contrast,
for the LCOE (Fig.~\ref{fig:min}b and d), there is a clear minimum for small
VRES gross shares on the side of the cheaper technology, in this example
figures, solar PV in AllCA and wind in MISO. Once surplus production begins,
leading to economically disadvantageous loss of value, the minimal LCOE region
starts to shift from lower installation and maintenance costs towards lower
surplus production. The effect is mitigated if alternative usages of the surplus
energy are found, however, unless more than half of the generation costs of the
surplus energy can be recovered in some way, at $100\,\%$ VRES gross share, the
LCOE-minimal mix still includes a significant share of the more expensive
technology.

These observations can be interpreted in two ways: First, they can be taken as
an indication that while in the beginning of the renewable build-up, least
generation costs pathways can be pursued without incurring additional backup
energy and subsequently, additional costs, the picture changes drastically as
soon as renewable penetrations reach beyond $30\,\%$-$50\,\%$. Then, surplus
production becomes an issue, technically as well as economically. One way of
tackling this challenge is to examine the backup energy-minimal wind/solar mix
and create a mixed renewable portfolio, even if generation costs alone clearly
favor only one technology.

Second, the situation can be viewed as a high incentive to make use of (and thus
gain from) VRES electricity excess generation. In the example of California with
a VRES gross share of $100\,\%$ and LCOE-minimizing mixes, the minimal LCOE if
surplus has no value is almost twice as high as in the case where all surplus
earns the same value as grid electricity, cf.~Fig.~\ref{fig:min}b. Surplus usage
can be achieved using inter-FERC-regional transmission, storage and demand-side
management, and coupling of the electricity system to heating and
transportation. A strong transmission grid that effectively allows for
long-range aggregation of wind generation is able to smooth it considerably
\citep{archer, holttinen05, sinden07, Widen:2011ys, Kempton:2010oq}, thus
providing a better match to the load. It has been shown in \cite{sarah,
sarahUS} that aggregation of load and generation shifts the backup-minimal mix
toward a higher wind share, and results in a reduction of backup energy needs by
about $20\,\%$ in the contiguous US. 

In wind-rich Ireland, a study has shown that wind integration can furthermore be
aided with flexible loads and hydro power plants, reporting a possible
surplus-free integration of $38\,\%$ wind into the Irish grid \citep{eirgrid}. 

Solar PV integration benefits much from short-term storage, which shifts the
backup-minimal mix towards solar PV \citep{morten}. An alternative to solar PV
combined with storage is concentrated solar power with inherent heat storage. 

Going beyond integration measures within the electricity sector, remaining
electrical excess generation can be used for heating or to produce
CO$_2$-neutral synthetic fuels for aviation and road transport.  This would lead
to a strong coupling of future energy infrastructures across the three big
energy sectors electricity, heating and cooling, and transportation.

\section*{Acknowledgments}

SB gratefully acknowledges financial support from O.\ and H.~St\"ocker as well
as M.\ and H.~Puschmann, BAF from a National Defense Science and Engineering
Graduate (NDSEG) fellowship, a National Science Foundation (NSF) graduate
fellowship, and a Stanford University Charles H. Leavell Graduate Student
Fellowship, and GBA from DONG Energy and the Danish Advanced Technology
Foundation.
The project underlying this report was supported by the German Federal Ministry
of Education and Research under grant no.~03SF0472C. The responsibility for the
contents lies with the authors.

\section*{References}

\bibliographystyle{model3-num-names}
\bibliography{literatur}

\begin{thebibliography}{30}
\providecommand{\natexlab}[1]{#1}
\providecommand{\url}[1]{\texttt{#1}}
\providecommand{\urlprefix}{URL }
\expandafter\ifx\csname urlstyle\endcsname\relax
  \providecommand{\doi}[1]{doi:\discretionary{}{}{}#1}\else
  \providecommand{\doi}{doi:\discretionary{}{}{}\begingroup
  \urlstyle{rm}\Url}\fi
\providecommand{\eprint}[2][]{\url{#2}}
\providecommand{\BIBand}{and}
\providecommand{\bibinfo}[2]{#2}
\ifx\xfnm\undefined \def\xfnm[#1]{\unskip,\space#1}\fi
\bibitem[{Budischak et~al.(2013)Budischak, Sewell, Thomson, Mach, Veron and
  Kempton}]{Budischak13}
\bibinfo{author}{Budischak\xfnm[ C.]}, \bibinfo{author}{Sewell\xfnm[ D.]},
  \bibinfo{author}{Thomson\xfnm[ H.]}, \bibinfo{author}{Mach\xfnm[ L.]},
  \bibinfo{author}{Veron\xfnm[ D.E.]}, \bibinfo{author}{Kempton\xfnm[ W.]}.
\newblock \bibinfo{title}{{Cost-minimized combinations of wind power, solar
  power and electrochemical storage, powering the grid up to 99.9\% of the
  time}}.
\newblock \bibinfo{journal}{Journal of Power Sources}
  \bibinfo{year}{2013};\bibinfo{volume}{225}(\bibinfo{number}{0}):\bibinfo{pages}{60
  -- 74}.
\newblock
  \bibinfo{note}{\url{http://dx.doi.org/10.1016/j.jpowsour.2012.09.054}}.
\bibitem[{Nelson et~al.(2012)Nelson, Johnston, Mileva, Fripp, Hoffman,
  Petros-Good et~al.}]{Nelson12}
\bibinfo{author}{Nelson\xfnm[ J.]}, \bibinfo{author}{Johnston\xfnm[ J.]},
  \bibinfo{author}{Mileva\xfnm[ A.]}, \bibinfo{author}{Fripp\xfnm[ M.]},
  \bibinfo{author}{Hoffman\xfnm[ I.]}, \bibinfo{author}{Petros-Good\xfnm[ A.]},
  et~al.
\newblock \bibinfo{title}{{High-resolution modeling of the western North
  American power system demonstrates low-cost and low-carbon futures}}.
\newblock \bibinfo{journal}{Energy Policy}
  \bibinfo{year}{2012};\bibinfo{volume}{43}(\bibinfo{number}{0}):\bibinfo{pages}{436
  -- 447}.
\newblock \bibinfo{note}{\url{http://dx.doi.org/10.1016/j.enpol.2012.01.031}}.
\bibitem[{Williams et~al.(2012)Williams, DeBenedictis, Ghanadan, Mahone, Moore,
  Morrow et~al.}]{Williams12}
\bibinfo{author}{Williams\xfnm[ J.H.]}, \bibinfo{author}{DeBenedictis\xfnm[
  A.]}, \bibinfo{author}{Ghanadan\xfnm[ R.]}, \bibinfo{author}{Mahone\xfnm[
  A.]}, \bibinfo{author}{Moore\xfnm[ J.]}, \bibinfo{author}{Morrow\xfnm[
  W.R.]}, et~al.
\newblock \bibinfo{title}{{The Technology Path to Deep Greenhouse Gas Emissions
  Cuts by 2050: The Pivotal Role of Electricity}}.
\newblock \bibinfo{journal}{Science}
  \bibinfo{year}{2012};\bibinfo{volume}{335}:\bibinfo{pages}{53--59}.
\newblock \bibinfo{note}{\url{http://dx.doi.org/10.1126/science.1208365}}.
\bibitem[{Hart and Jacobson(2011)}]{hart11}
\bibinfo{author}{Hart\xfnm[ E.K.]}, \bibinfo{author}{Jacobson\xfnm[ M.Z.]}.
\newblock \bibinfo{title}{{A Monte Carlo approach to generator portfolio
  planning and carbon emissions assessments of systems with large penetrations
  of variable renewables}}.
\newblock \bibinfo{journal}{Renewable Energy}
  \bibinfo{year}{2011};\bibinfo{volume}{36}(\bibinfo{number}{8}):\bibinfo{pages}{2278
  -- 2286}.
\newblock \bibinfo{note}{\url{http://dx.doi.org/10.1016/j.renene.2011.01.015}}.
\bibitem[{Hand et~al.(2012)Hand, Baldwin, DeMeo, Reilly, Mai, Arent
  et~al.}]{NREL_re_futures}
\bibinfo{author}{Hand\xfnm[ M.]}, \bibinfo{author}{Baldwin\xfnm[ S.]},
  \bibinfo{author}{DeMeo\xfnm[ E.]}, \bibinfo{author}{Reilly\xfnm[ J.]},
  \bibinfo{author}{Mai\xfnm[ T.]}, \bibinfo{author}{Arent\xfnm[ D.]}, et~al.
\newblock \bibinfo{title}{{Renewable Electricity Futures Study}}.
\newblock
  \bibinfo{howpublished}{\url{http://www.nrel.gov/analysis/re_futures/}};
  \bibinfo{year}{2012}.
\newblock \bibinfo{note}{Eds. 4 vols. NREL/TP-6A20-52409}.
\bibitem[{{McKinsey \& Company} et~al.(2010){McKinsey \& Company}, {KEMA}, {The
  Energy Futures Lab at Imperial College London}, {Oxford Economics} and
  {ECF}}]{ecf2050}
\bibinfo{author}{{McKinsey \& Company}\xfnm[]},
  \bibinfo{author}{{KEMA}\xfnm[]}, \bibinfo{author}{{The Energy Futures Lab at
  Imperial College London}\xfnm[]}, \bibinfo{author}{{Oxford
  Economics}\xfnm[]}, \bibinfo{author}{{ECF}\xfnm[]}.
\newblock \bibinfo{title}{{Roadmap 2050 -- A practical guide to a prosperous,
  low-carbon Europe}}.
\newblock \bibinfo{type}{Tech. Rep.}; {European Climate Foundation};
  \bibinfo{address}{\url{http://www.roadmap2050.eu/}}; \bibinfo{year}{2010}.
\newblock \bibinfo{note}{{Online, accessed June 2012}}.
\bibitem[{{F\"ursch} et~al.(2011){F\"ursch}, {Hagspiel}, {J\"agemann}, {Nagl},
  {Lindenberger}, {Glotzbach} et~al.}]{energynautics}
\bibinfo{author}{{F\"ursch}\xfnm[ M.]}, \bibinfo{author}{{Hagspiel}\xfnm[ S.]},
  \bibinfo{author}{{J\"agemann}\xfnm[ C.]}, \bibinfo{author}{{Nagl}\xfnm[ S.]},
  \bibinfo{author}{{Lindenberger}\xfnm[ D.]},
  \bibinfo{author}{{Glotzbach}\xfnm[ L.]}, et~al.
\newblock \bibinfo{title}{{Roadmap 2050 -- a closer look}}.
\newblock \bibinfo{type}{Tech. Rep.}; energynautics and {Institute for Energy
  Economy at the University of Cologne};
  \bibinfo{address}{\url{http://www.energynautics.com/news/}};
  \bibinfo{year}{2011}.
\bibitem[{Gelman(2013)}]{DOE12}
\bibinfo{author}{Gelman\xfnm[ R.]}.
\newblock \bibinfo{title}{{2012 Renewable Energy Data Book}}.
\newblock \bibinfo{type}{Tech. Rep.}; {US Department of Energy};
  \bibinfo{year}{2013}.
\newblock \bibinfo{note}{Online available at
  \url{http://www.nrel.gov/news/press/2013/5302.html}, retrieved Sep 2014}.
\bibitem[{Jacobson and Delucchi(2011)}]{jacobson11}
\bibinfo{author}{Jacobson\xfnm[ M.Z.]}, \bibinfo{author}{Delucchi\xfnm[ M.A.]}.
\newblock \bibinfo{title}{{Providing all global energy with wind, water, and
  solar power, Part I: Technologies, energy resources, quantities and areas of
  infrastructure, and materials}}.
\newblock \bibinfo{journal}{Energy Policy}
  \bibinfo{year}{2011};\bibinfo{volume}{39}(\bibinfo{number}{3}):\bibinfo{pages}{1154–1169}.
\newblock \bibinfo{note}{\url{http://dx.doi.org/10.1016/j.enpol.2010.11.040}}.
\bibitem[{Delucchi and Jacobson(2011)}]{delucchi11}
\bibinfo{author}{Delucchi\xfnm[ M.A.]}, \bibinfo{author}{Jacobson\xfnm[ M.Z.]}.
\newblock \bibinfo{title}{{Providing all global energy with wind, water, and
  solar power, Part II: Reliability, System and Transmission Costs, and
  Policies}}.
\newblock \bibinfo{journal}{Energy Policy}
  \bibinfo{year}{2011};\bibinfo{volume}{39}(\bibinfo{number}{3}):\bibinfo{pages}{1170--1190}.
\newblock \bibinfo{note}{\url{http://dx.doi.org/10.1016/j.enpol.2010.11.045}}.
\bibitem[{{Becker} et~al.(2014{\natexlab{a}}){Becker}, {Frew}, {Andresen},
  {Zeyer}, {Schramm}, {Greiner} et~al.}]{sarahUS}
\bibinfo{author}{{Becker}\xfnm[ S.]}, \bibinfo{author}{{Frew}\xfnm[ B.A.]},
  \bibinfo{author}{{Andresen}\xfnm[ G.B.]}, \bibinfo{author}{{Zeyer}\xfnm[
  T.]}, \bibinfo{author}{{Schramm}\xfnm[ S.]}, \bibinfo{author}{{Greiner}\xfnm[
  M.]}, et~al.
\newblock \bibinfo{title}{{Features of a fully renewable US electricity system:
  Optimal mixes of wind and solar PV and transmission grid extensions}}.
\newblock \bibinfo{journal}{Energy}
  \bibinfo{year}{2014}{\natexlab{a}};\bibinfo{volume}{72}:\bibinfo{pages}{443–458}.
\newblock \bibinfo{note}{Online at
  \url{http://dx.doi.org/10.1016/j.energy.2014.05.067}; preprint available at
  \url{http://arxiv.org/abs/1402.2833}}.
\bibitem[{{Corcoran} et~al.(2012){Corcoran}, {Jenkins} and
  {Jacobson}}]{bethany}
\bibinfo{author}{{Corcoran}\xfnm[ B.A.]}, \bibinfo{author}{{Jenkins}\xfnm[
  N.]}, \bibinfo{author}{{Jacobson}\xfnm[ M.Z.]}.
\newblock \bibinfo{title}{{Effects of aggregating electric load in the United
  States}}.
\newblock \bibinfo{journal}{Energy Policy}
  \bibinfo{year}{2012};\bibinfo{volume}{46}:\bibinfo{pages}{399--416}.
\newblock \bibinfo{note}{\url{http://dx.doi.org/10.1016/j.enpol.2012.03.079}}.
\bibitem[{{Saha et\,al}(2010)}]{saha}
\bibinfo{author}{{Saha et\,al}\xfnm[ S.]}.
\newblock \bibinfo{title}{{The NCEP Climate Forecast System Reanalysis}}.
\newblock \bibinfo{journal}{{Bull Amer Meteor Soc}}
  \bibinfo{year}{2010};\bibinfo{volume}{91}(\bibinfo{number}{8}):\bibinfo{pages}{1015--1057}.
\newblock \bibinfo{note}{\url{http://dx.doi.org/10.1175/2010BAMS3001.1}}.
\bibitem[{{Andresen} et~al.(2014){Andresen}, {S{\o}ndergaard} and
  {Greiner}}]{REAtlas}
\bibinfo{author}{{Andresen}\xfnm[ G.B.]},
  \bibinfo{author}{{S{\o}ndergaard}\xfnm[ A.A.]},
  \bibinfo{author}{{Greiner}\xfnm[ M.]}.
\newblock \bibinfo{title}{{Validation of Danish wind time series from a new
  global renewable energy atlas for energy system analysis}}.
\newblock \bibinfo{journal}{Submitted for review}
  \bibinfo{year}{2014};\bibinfo{note}{Preprint available at
  \url{http://arxiv.org/abs/1409.3353}}.
\bibitem[{{S{\o}ndergaard}(2013)}]{anders}
\bibinfo{author}{{S{\o}ndergaard}\xfnm[ A.A.]}.
\newblock \bibinfo{title}{{Development of a Renewable Energy Atlas and Extreme
  Event Analysis in Renewable Energy Systems}}.
\newblock \bibinfo{type}{{M.~Sc. thesis}}; Aarhus University;
  \bibinfo{address}{Denmark}; \bibinfo{year}{2013}.
\bibitem[{{C.~Potter and B.~Nijssen}(2009)}]{NREL_wwind}
\bibinfo{author}{{C.~Potter and B.~Nijssen}\xfnm[]}.
\newblock \bibinfo{title}{{Development of Regional Wind Resource and Wind Plant
  Output Datasets}}.
\newblock \bibinfo{type}{Tech. Rep.}; 3TIER, for NREL, under supervision of
  D.~Lew; \bibinfo{address}{Seattle, Washington}; \bibinfo{year}{2009}.
\newblock \bibinfo{note}{Final data sets available at
  \url{http://www.nrel.gov/electricity/transmission/western_wind_dataset.html}}.
\bibitem[{{M.~Brower}(2009)}]{NREL_ewind}
\bibinfo{author}{{M.~Brower}\xfnm[]}.
\newblock \bibinfo{title}{{Development of Eastern Regional Wind Resource and
  Wind Plant Output Datasets}}.
\newblock \bibinfo{type}{Tech. Rep.}; AWS Truewind LLC, for NREL, under
  supervision of D.~Corbus; \bibinfo{address}{Albany, New York};
  \bibinfo{year}{2009}.
\newblock \bibinfo{note}{Final data sets available at
  \url{http://www.nrel.gov/electricity/transmission/eastern_wind_dataset.html}}.
\bibitem[{Archer and Jacobson(2007)}]{archer}
\bibinfo{author}{Archer\xfnm[ C.L.]}, \bibinfo{author}{Jacobson\xfnm[ M.Z.]}.
\newblock \bibinfo{title}{{Supplying Baseload Power and Reducing Transmission
  Requirements by Interconnecting Wind Farms}}.
\newblock \bibinfo{journal}{{Journal of Applied Meteorology and Climatology}}
  \bibinfo{year}{2007};\bibinfo{volume}{46}(\bibinfo{number}{11}):\bibinfo{pages}{1701--1717}.
\newblock \bibinfo{note}{\url{http://dx.doi.org/10.1175/2007JAMC1538.1}}.
\bibitem[{Holttinen(2005)}]{holttinen05}
\bibinfo{author}{Holttinen\xfnm[ H.]}.
\newblock \bibinfo{title}{{Impact of hourly wind power variations on the system
  operation in the Nordic countries}}.
\newblock \bibinfo{journal}{Wind Energy}
  \bibinfo{year}{2005};\bibinfo{volume}{8}(\bibinfo{number}{2}):\bibinfo{pages}{197--218}.
\newblock \bibinfo{note}{\url{http://dx.doi.org/10.1002/we.143}}.
\bibitem[{Sinden(2007)}]{sinden07}
\bibinfo{author}{Sinden\xfnm[ G.]}.
\newblock \bibinfo{title}{{Characteristics of the UK wind resource: Long-term
  patterns and relationship to electricity demand}}.
\newblock \bibinfo{journal}{Energy Policy}
  \bibinfo{year}{2007};\bibinfo{volume}{35}(\bibinfo{number}{1}):\bibinfo{pages}{112–127}.
\newblock \bibinfo{note}{\url{http://dx.doi.org/10.1016/j.enpol.2005.10.003}}.
\bibitem[{Wiemken et~al.(2001)Wiemken, Beyer, Heydenreich and
  Kiefer}]{wiemken01}
\bibinfo{author}{Wiemken\xfnm[ E.]}, \bibinfo{author}{Beyer\xfnm[ H.]},
  \bibinfo{author}{Heydenreich\xfnm[ W.]}, \bibinfo{author}{Kiefer\xfnm[ K.]}.
\newblock \bibinfo{title}{{Power characteristics of PV ensembles: experiences
  from the combined power production of 100 grid connected PV systems
  distributed over the area of Germany}}.
\newblock \bibinfo{journal}{Solar Energy}
  \bibinfo{year}{2001};\bibinfo{volume}{70}(\bibinfo{number}{6}):\bibinfo{pages}{513–518}.
\newblock
  \bibinfo{note}{\url{http://dx.doi.org/10.1016/S0038-092X(00)00146-8}}.
\bibitem[{Mills and Wiser(2010)}]{mills10}
\bibinfo{author}{Mills\xfnm[ A.]}, \bibinfo{author}{Wiser\xfnm[ R.]}.
\newblock \bibinfo{title}{{Implications of Wide-Area Geographic Diversity for
  Short- Term Variability of Solar Power}}.
\newblock \bibinfo{type}{Tech. Rep.}; Lawrence Berkeley National Laboratory;
  \bibinfo{year}{2010}.
\bibitem[{Wid\'{e}n(2011)}]{Widen:2011ys}
\bibinfo{author}{Wid\'{e}n\xfnm[ J.]}.
\newblock \bibinfo{title}{{Correlations between large-scale solar and wind
  power in a future scenario for Sweden}}.
\newblock \bibinfo{journal}{IEEE Transactions on Sustainable Energy}
  \bibinfo{year}{2011};\bibinfo{volume}{2}(\bibinfo{number}{2}):\bibinfo{pages}{177--184}.
\newblock \bibinfo{note}{\url{http://dx.doi.org/10.1109/TSTE.2010.2101620}}.
\bibitem[{Kempton et~al.(2010)Kempton, Pimenta, Veron and
  Colle}]{Kempton:2010oq}
\bibinfo{author}{Kempton\xfnm[ W.]}, \bibinfo{author}{Pimenta\xfnm[ F.M.]},
  \bibinfo{author}{Veron\xfnm[ D.E.]}, \bibinfo{author}{Colle\xfnm[ B.A.]}.
\newblock \bibinfo{title}{Electric power from offshore wind via synoptic-scale
  interconnection}.
\newblock \bibinfo{journal}{Proceedings of the National Academy of Sciences}
  \bibinfo{year}{2010};\bibinfo{volume}{107}(\bibinfo{number}{16}):\bibinfo{pages}{7240--7245}.
\newblock \bibinfo{note}{\url{http://dx.doi.org/10.1073/pnas.0909075107}}.
\bibitem[{{Energy and Environmental Economics}(2012)}]{E3report}
\bibinfo{author}{{Energy and Environmental Economics}\xfnm[]}.
\newblock \bibinfo{title}{{Cost and Performance Review of Generation
  Technologies: Recommendations for WECC 10- and 20-Year Studies}}.
\newblock \bibinfo{type}{Tech. Rep.}; E3, San Francisco, CA;
  \bibinfo{year}{2012}.
\newblock \bibinfo{note}{{Online available at
  \url{http://www.nwcouncil.org/media/6867814/E3_GenCapCostReport_finaldraft.pdf},
  retrieved Aug 2013}}.
\bibitem[{{US Army Corps of Engineers}(2011)}]{cwccis}
\bibinfo{author}{{US Army Corps of Engineers}\xfnm[]}.
\newblock \bibinfo{title}{{Civil Works Construction Cost Index System}}.
\newblock \bibinfo{type}{Tech. Rep.}; {US Army Corps of Engineers};
  \bibinfo{address}{\url{http://planning.usace.army.mil/toolbox/library/EMs/em1110.2.1304.pdf}};
  \bibinfo{year}{2011}.
\newblock \bibinfo{note}{{Online, accessed Aug 2013}}.
\bibitem[{Lazard(2014)}]{Lazard2014}
\bibinfo{author}{Lazard\xfnm[]}.
\newblock \bibinfo{title}{{Lazard's Levelized Cost Of Energy Analysis--Version
  8.0}}.
\newblock \bibinfo{year}{2014}.
\bibitem[{{Becker} et~al.(2014{\natexlab{b}}){Becker}, {Rodr\'iguez},
  {Andresen}, {Schramm} and {Greiner}}]{sarah}
\bibinfo{author}{{Becker}\xfnm[ S.]}, \bibinfo{author}{{Rodr\'iguez}\xfnm[
  R.A.]}, \bibinfo{author}{{Andresen}\xfnm[ G.B.]},
  \bibinfo{author}{{Schramm}\xfnm[ S.]}, \bibinfo{author}{{Greiner}\xfnm[ M.]}.
\newblock \bibinfo{title}{{Transmission grid extensions during the build-up of
  a fully renewable pan-European electricity supply}}.
\newblock \bibinfo{journal}{Energy}
  \bibinfo{year}{2014}{\natexlab{b}};\bibinfo{volume}{64}:\bibinfo{pages}{404--418}.
\newblock \bibinfo{note}{\url{http://dx.doi.org/10.1016/j.energy.2013.10.010};
  preprint available at \url{http://arxiv.org/abs/1307.1723}}.
\bibitem[{EirGrid(2008)}]{eirgrid}
\bibinfo{author}{EirGrid\xfnm[]}.
\newblock \bibinfo{title}{{All Island Grid Study}}.
\newblock \bibinfo{year}{2008}.
\newblock \bibinfo{note}{Online available at
  \url{http://www.eirgrid.com/renewables/all-islandgridstudy/}, retrieved Sep
  2014}.
\bibitem[{{Rasmussen} et~al.(2012){Rasmussen}, {Andresen} and
  {Greiner}}]{morten}
\bibinfo{author}{{Rasmussen}\xfnm[ M.G.]}, \bibinfo{author}{{Andresen}\xfnm[
  G.B.]}, \bibinfo{author}{{Greiner}\xfnm[ M.]}.
\newblock \bibinfo{title}{{Storage and balancing synergies in a fully or highly
  renewable pan-European power system}}.
\newblock \bibinfo{journal}{Energy Policy}
  \bibinfo{year}{2012};\bibinfo{volume}{51}:\bibinfo{eid}{642-651}.
\newblock \bibinfo{note}{\url{http://dx.doi.org/10.1016/j.enpol.2012.09.009}}.

\end{thebibliography}

\end{document}